\documentstyle[psfig]{mn2e}
\title{Weak Gravitational Lensing by Dark Clusters}
\author[N. N. Weinberg and M. Kamionkowski]{Nevin N. Weinberg and
Marc Kamionkowski \\
California Institute of Technology, Mail Code 130-33, Pasadena, CA 91125 USA}

\onecolumn
\pagerange{\pageref{firstpage}-\pageref{lastpage}}
\date{March 2002}
\pubyear{2002}

\begin{document}

\maketitle

\label{firstpage}

\begin{abstract}
We calculate the abundance of dark-matter concentrations that
are sufficiently overdense to produce a detectable
weak-gravitational-lensing signal.  Most of these
overdensities are virialized halos containing identifiable
X-ray and/or optical clusters.  However, a significant fraction
are nonvirialized, cluster-mass, overdensities still in the 
process of gravitational
collapse---these should produce significantly weaker or no
X-ray emission.  Our predicted abundance of such dark clusters
are consistent with the abundance implied by the Erben et
al. (2000) detection of an apparent dark lens.  Weak lensing by these
nonvirialized objects will need to be considered when
determining cosmological parameters with the lens abundance in
future weak-lensing surveys.  Such weak lenses should also help
shed light on the process of cluster formation.
\end{abstract}



\section{INTRODUCTION}

Weak gravitational lensing due to the deep gravitational
potential of a galaxy cluster gives rise to a detectable weak
distortion of the images of background galaxies.  This weak
shear has now been detected around roughly 30 clusters and been
used to map the total dark-matter mass in the clusters as well as
the dark-matter distributions within the clusters (see Bartelmann
\& Schneider 2001, Mellier 1999). Weak lensing also has the potential
to map the mass distribution on even
larger scales \cite{Mir91,Bla91,Kai92,BarSch92,Ste97,Kametal98}.  Just
last year, four groups independently reported detection of cosmic
shear, distortions to background galaxies induced by weak
gravitational lensing by mass inhomogeneities on {\it few}-Mpc scales
along the line of sight
\cite{BacRefEll00,KaiWilLup00,Witetal00,vanetal00}.  It is
apparent that in the future, such
cosmic-shear surveys will have the sensitivity to identify
galaxy clusters in the field.  Since such surveys will probe
the {\it total} mass directly, it could provide a powerful new
technique for determining the cluster-halo abundance and thus the
power-spectrum amplitude $\sigma_8$ and matter density
$\Omega_m$ (e.g., Kruse \& Schneider 1999; Reblinsky et al. 1999).

In fact, one spectroscopically-confirmed cluster has already
been detected via its gravitational-lensing effect on background
galaxies \cite{Witetal01}.  More intriguing is the apparent dark
lens discovered by Erben et al. (2000).  This lensing signal
corresponds to a $\sim 10^{14} M_{\odot}$ mass concentration, but there
is no obvious corresponding galaxy overdensity (Gray et al. 2001) and only
faint (if any) X-ray emission. Evidence for other apparent dark lenses has
been reported by Miralles et al. (2002) and Koopmans et al. (2000),
the latter involving a detection through strong, rather than weak,
lensing.

In retrospect, the existence of such dark concentrations should
not come as too much of a surprise.  Galaxy clusters form
at rare (e.g., $>3\sigma$) high-density peaks of a Gaussian
primordial distribution.  Thus, for every virialized cluster,
there should be a significant number of proto-clusters (e.g.,
$2\sigma-3\sigma$ peaks), mass
overdensities that have not yet undergone gravitational collapse
and virialized, but which have begun to break away from the
cosmological expansion. The timescale for collapse of
cluster-mass objects is large, and the overdensities can be very
large even before they have virialized.  It should thus not be too
surprising if such objects produce a weak-lensing signal that
resembles that from virialized clusters.  

These proto-clusters should contain galaxies and maybe a few
groups that later merge to form the cluster.  Since the X-ray
luminosity is a very rapidly varying function of the virialized
mass, the summed X-ray emission from these objects should be
much smaller than that from a fully virialized cluster of the
same mass.  When we refer to these proto-clusters as ``dark,''
we thus mean that they should be X-ray underluminous.
Strictly speaking, the mass-to-light ratios of these clusters
should be comparable to those for ordinary clusters.  However,
high-redshift clusters may be difficult to pick out in galaxy
surveys, and these proto-clusters should have a sky
density a few times smaller.  Thus it would not be surprising
if these dark lenses had no readily apparent corresponding
galaxy overdensity.

In this paper we calculate the abundance of dark and virialized
lenses.  To do so, we first determine the overdensity required to
produce a detectable weak-lensing signal as a function of redshift. We
consider several different density profiles including a homogeneous
sphere, an isothermal sphere, a Navarro, Frenk, \& White
\cite{NFW97,NFW96,NFW95} profile and a Hernquist \cite{Her90} profile.
We then use the spherical-top-hat-collapse (STHC) model to determine
the differential abundance of overdensities as a function of position
along their evolutionary cycle. Using the aperture mass technique (Schneider
1996) we can then determine the sky density and redshift distribution of 
halos that are sufficiently overdense to produce a detectable weak-lensing 
signal.

As our results below will show, there should be roughly one dark lens
for every 5--10 virialized lenses discovered by weak lensing. It is
worthwhile to point out that this result is robust in that the ratio of  
dark to virialized lenses is not expected to be very sensitive to the
amount of {\it observational} noise in the lensing map, i.e., observational
noise will equally affect the detectability of both types of lenses.
Therefore, although our results are obtained by assuming the only source
of noise is the intrinsic ellipticity distribution of the source galaxies --
in accordance with other such theoretical weak-lensing studies
found in the literature -- the predicted relative abundance of dark and
virialized lenses will not change very much if we made a more exact 
estimate of the total noise in weak-lensing maps. It is also encouraging
to note that given the sky coverage and average image size of 
weak-lensing maps to date, the number of dark lenses we would expect to 
have seen is of order unity
and therefore consistent with the detection (Erben et al. 2000,
Miralles et al. 2002) of one or two dark lenses.

\section{Minimum overdensity required to produce weak-lensing signal}

In this Section we provide the conditions for an overdensity of
mass $M$ and radius $R$ at redshift $z$ to produce a detectable
weak-lensing signal.  Following the procedure of Bartelmann \&
Schneider (2001) (see also Schneider 1996, Seitz \& Schneider
1997, Kruse \& Schneider 1999)
we determine the dependence of a lensing system's signal-to-noise
ratio on that system's overdensity and redshift.

In a weak-lensing map, a mass overdensity causes the image of the
background source galaxies to be tangentially sheared. Noise is
introduced by both the intrinsic ellipticity of these background
galaxies as well as by the presence of foreground galaxies in the
image. To arrive at a signal-to-noise relation for a weak-lensing
system, consider $N$ galaxy images each at angular position
$\mbox{\boldmath$\theta$\unboldmath}_i=(\theta_i \cos \phi_i, \theta_i
\sin \phi_i)$ with tangential ellipticity
$\epsilon_t(\mbox{\boldmath$\theta$\unboldmath}_i)$ and within
a lens-centered annulus that is bounded by
angular radii $\theta_{\rm{in}} \leq \theta_i \leq
\theta_{\rm{out}}$. The shear $\gamma$ is related linearly to the
dimensionless surface mass density of the lens, which is the physical
surface mass density $\Sigma(\mbox{\boldmath$\theta$\unboldmath})$
divided by the critical surface mass density $\Sigma_{\rm{crit}}$.
For a lens at redshift $z_d$ and a source at redshift $z_s$,
\begin{equation}
\Sigma_{\rm{crit}}(z_d;z_s)=\frac{c^2}{4 \pi G}\frac{D_s}{D_d D_{ds}},
\end{equation}
where $D_d$, $D_s$ and $D_{ds}$ are the angular-diameter distances
between the lens and the observer, the source galaxy and
the observer, and the lens and the source, respectively.
To account for the redshift distribution of the source
galaxies, define (Seitz \& Schneider 1997)
\begin{equation}
Z(z_s; z_d) \equiv
\frac{\lim_{z_s \rightarrow \infty}
\Sigma_{\rm{crit}}(z_d;z_s)}{\Sigma_{\rm{crit}}(z_d;z_s)}=
\frac{\Sigma_{\rm{crit}_{\infty}}(z_d)}
{\Sigma_{\rm{crit}}(z_d;z_s)}.
\end{equation}
Then the dimensionless surface mass density is given
by,
\begin{equation}
\kappa(\mbox{\boldmath$\theta$\unboldmath},z_s)=
\frac{\Sigma(\mbox{\boldmath$\theta$\unboldmath})}{\Sigma_{\rm
{crit}}}=\frac{\Sigma(\mbox{\boldmath$\theta$\unboldmath})}{\Sigma_{\rm
{crit}_{\infty}}} \frac{\Sigma_{\rm {crit}_{\infty}}}{\Sigma_{\rm
{crit}}}
\equiv \kappa(\mbox{\boldmath$\theta$\unboldmath})Z(z_s; z_d).
\end{equation}
Furthermore, the linear relation between the shear and surface mass
density implies that they have the same dependence on source redshift
so that $\gamma(\mbox{\boldmath$\theta$\unboldmath},z_s) \equiv
Z(z_s;z_d)\gamma(\mbox{\boldmath$\theta$\unboldmath})$. For the rest of
this paper any reference to $\kappa$ or $\gamma$ refers to
$\kappa(\mbox{\boldmath$\theta$\unboldmath})$ and
$\gamma(\mbox{\boldmath$\theta$\unboldmath})$, respectively.
Assuming the intrinsic orientation of galaxy sources is random, the expectation
value of the image ellipticity is \cite{SeiSch97,BarSch01}
\begin{equation}
E(\epsilon) \approx \langle
Z \rangle \gamma(\mbox{\boldmath$\theta$\unboldmath}),
\end{equation}
where
\begin{equation}
\langle Z \rangle = \int dz_s \; p_z(z_s) Z(z_s;z_d),
\end{equation}
and $p_z(z_s)$ is the redshift distribution of source galaxies.
The function $\langle Z \rangle = \langle Z \rangle
(z_d)$---of order unity for the redshifts considered---allows a
source redshift distribution to be collapsed onto a single
redshift $z_s$ satisfying $Z(z_s)= \langle Z \rangle$ (see
Bartelmann \& Schneider 2001).

Using the $M_{\rm{ap}}$-statistics introduced by Schneider (1996), define a
discretized estimator for the spatially filtered mass inside a circular
aperture of angular radius $\theta$,
\begin{equation}
M_{\rm{ap}} \equiv \frac{1}{n}\sum_{i=1}^N \epsilon_t(\mbox{\boldmath$\theta$\unboldmath}_i)
\, Q(|\mbox{\boldmath$\theta$\unboldmath}_i|),
\end{equation}
where $n$ is the number density of galaxy images and $Q$
is a weight function that will be chosen
later so as to maximize the signal-to-noise ratio of the
estimator. Assuming the ellipticities of different images are uncorrelated
the dispersion of $M_{\rm{ap}}$ can be obtained by squaring (6) and taking
the expectation value, yielding
\begin{equation}
\sigma^2 = \frac{\sigma_{\epsilon}^2}{2n^2}
\sum_{i=1}^N Q^2(|\mbox{\boldmath$\theta$\unboldmath}_i|),
\end{equation}
where $\sigma_{\epsilon}$ is the dispersion of the two component
ellipticity. By (4) the expectation value of $M_{\rm{ap}}$ is,
\begin{equation}
\langle M_{\rm{ap}}\rangle = \frac{\langle Z \rangle}{n}\sum_{i=1}^N
\gamma_t(\mbox{\boldmath$\theta$\unboldmath}_i)
\, Q(|\mbox{\boldmath$\theta$\unboldmath}_i|),
\end{equation}
where $\gamma_t$ is the tangential shear. Taking the ensemble average of (8) over the
probability distribution for the galaxy positions gives,
\begin{equation}
\langle M_{\rm{ap}}\rangle_c = 2\pi\langle Z \rangle
\int_{\theta_{\rm{in}}}^{\theta_{\rm{out}}}
d\theta \, \theta \, \langle \gamma_t \rangle (\theta) \, Q(\theta),
\end{equation}
where $ \langle \gamma_t \rangle (\theta)$ is the mean tangential shear on a circle of
angular radius $\theta$ and the subscript `c' stands for continuous. Similarly, we can
take the ensemble average of the dispersion (7), to obtain
\begin{equation}
\sigma^2_c = \frac{\pi \sigma_{\epsilon}^2}{n}
\int_{\theta_{\rm{in}}}^{\theta_{\rm{out}}}
d\theta \, \theta \, Q^2(\theta).
\end{equation}
The ensemble-averaged signal-to-noise ratio is then,
\begin{equation}
\frac{S}{N}=\frac{\langle M_{\rm{ap}}\rangle_c}{\sigma_c}
=\frac{2 \langle Z \rangle \sqrt{\pi n}}{\sigma_{\epsilon}}
\frac{\int_{\theta_{\rm{in}}}^{\theta_{\rm{out}}}
d\theta \, \theta \langle \gamma_t \rangle (\theta) \, Q(\theta)}
{\sqrt{\int_{\theta_{\rm{in}}}^{\theta_{\rm{out}}}
d\theta \, \theta \, Q^2(\theta)}}.
\end{equation}
By the Cauchy-Schwarz inequality the signal-to-noise ratio of the
estimator is maximized if
\begin{equation}
Q(\theta) \propto  \langle \gamma_t \rangle(\theta).
\end{equation}
Since
\begin{equation}
\langle \gamma_t \rangle(\theta)=
\bar{\kappa}(\theta)-\langle \kappa \rangle(\theta) ,
\end{equation}
\cite{Bar95} where $\langle \kappa \rangle(\theta)  $ is the
dimensionless mean surface mass density on a circle of radius $\theta$
and $\bar{\kappa}(\theta)$ is the dimensionless mean surface mass
density within a circle of radius $\theta$, the maximized
signal-to-noise ratio becomes
\begin{equation} \frac{S}{N}=\frac{2
\langle Z \rangle \sqrt{\pi n}}{\sigma_{\epsilon}}
\sqrt{\int_{\theta_{\rm{in}}}^{\theta_{\rm{out}}} d\theta \, \theta \,
[\bar{\kappa}(\theta)-\langle \kappa \rangle(\theta) ]^2}.
\end{equation}
If instead of using a maximized weight function $Q$ we chose one of the often used
generic weight functions given in Schneider et al. (1998), our estimate of
the signal-to-noise ratio for a given lens would be slightly smaller. In an upcoming paper
(Weinberg \& Kamionkowski 2002) we show that although using such a weight function reduces 
the predicted abundance of dark and virialized lenses somewhat, the principle result of this
paper, namely that the relative abundance of dark to virialized lenses is 10--20\%,
is virtually unchanged so long as $\theta_{\rm{out}} \ga 3$ arcmin.

To compute the signal-to-noise ratio for a lens with a given density
profile we need to determine the mean tangential shear of the source
galaxies. Different density profiles will in general produce different
shear patterns. In particular, the more cuspy a profile is the stronger its lensing
signal. Of course this becomes more complicated when considering
profiles with power-law breaks. For instance, although the NFW
profile goes as $r^{-1}$ at small radii while the isothermal sphere
goes as $r^{-2}$, at larger radii the former varies as $r^{-3}$ while
the latter remains at $r^{-2}$. The net effect, as we will show, is
that the NFW profile yields a stronger signal compared to the
isothermal sphere for lenses at reasonable redshifts. That said, we
consider a variety of profiles to account for the full range of
possibilities and to study the dependence of our results on these
profiles. Specifically, we compare the calculated abundances assuming
the overdensity is a point mass, a uniform-density sphere, an
isothermal sphere, an NFW profile and a Hernquist profile. For an object of
a given mass, mean overdensity, and density profile, we can solve
for the parameters of the given profile (e.g., the radius, the velocity dispersion,
the scale radius, the scale density, etc.) and determine, using equation (14), whether
such an object produces a sufficiently large weak-lensing signal-to-noise ratio so
as to be detectable. Note that for an overdensity with angular radius smaller than
the size of the lensing image, the shear pattern beyond the radius of the overdensity
will be that of a point mass. Furthermore, if the angular radius is larger than the
image size then the lensing signal is determined by just the mass
$M_P$ within the projected image radius $P=\theta_{\rm{out}}D_d$ and
not the mass outside this radius. The derivation of the
signal-to-noise relation for each of these profiles is given in the
Appendix.

To produce a detectable signal, an overdensity must be large
enough to yield a signal-to-noise ratio greater than some minimum
value. For the calculations done in this paper we adopt the following
fiducial values, unless stated otherwise: $(S/N)_{\rm{min}}=5$,
$\theta_{\rm{out}} = 5$ arcmin, the number density of galaxy images is
$n=30$ arcmin$^{-2}$, and $\sigma_{\epsilon} = 0.2$. The minimum
nonlinear overdensity corresponds to a particular position along the
linear-theory evolutionary cycle. In the next Section we discuss how
we relate the minimum \emph{nonlinear} overdensity to a corresponding
minimum \emph{linear}-theory overdensity. This will enable us to apply
the Press-Schechter formalism to obtain an estimate of the abundance of
overdensities that produce a weak-lensing signal as a function of
redshift.

\section{Dynamics}

We use the STHC model to relate the minimum non-linear
overdensity needed for a detectable weak-lensing signal at a given
redshift to a minimum linear-theory overdensity. According to STHC
the non-linear evolution of cosmic density fluctuations is
approximated by a dynamical model in which the initial linear
perturbation is an isolated, uniform sphere surrounded by unperturbed
matter. Gravitational instability causes the initially small linear
perturbation to grow and enter the nonlinear regime, ultimately
forming a virialized object that is decoupled from the cosmological
background. In order to avoid the collapse to infinite density
predicted by the solution of STHC, we invoke a simple smoothing scheme
that allows us to map a linear overdensity greater than the critical
linear density contrast, $\delta_c \sim 1.69$, to a finite nonlinear
overdensity. In what follows we shall consider the STHC model in a
$\Lambda$CDM universe. Following the derivation of the relevant STHC
formula, we present our smoothing scheme.
Finally, we discuss how we distinguish ``virialized'' clusters from
those that have not yet collapsed.

For a flat cosmology with a cosmological constant, the change in the
proper radius, $r$, with scale factor $a$ for a uniform spherical
overdensity of fixed mass $M$ is given by (see Peebles 1984, Eke et
al. 1996)
\begin{equation}
\left( \frac{dr}{da} \right)^2 =
   \frac{r^{-1} + \omega r^2 -\beta}{a^{-1}+\omega a^2},
\end{equation}
where $a=(1+z)^{-1}$, $\beta$ is a constant which is positive for
overdensities and
\begin{equation}
\omega=(\Omega_0^{-1}-1),
\end{equation}
where $\Omega_0$ is the cosmological density parameter.
Note that the units of $r$ are such that $(3M/4\pi\rho_0)^{1/3} \equiv
1$ where $\rho_0$ is the cosmological background density at $z=0$.
Separating the variables in equation (15) and integrating gives
\begin{equation}
\int_0^r \frac{r'^{1/2}}{\left(\omega r'^3 -\beta r'
+1 \right)^{1/2}}dr'=\int_0^a \frac{a'^{1/2}}{\left(\omega a'^3+1\right)^{1/2}}da'.
\end{equation}
Solving for the root of the numerator in equation (15) gives the turnaround
radius (i.e., radius at maximum expansion), $r_{ta}$, as a function of
the density parameter $\omega$ and perturbation amplitude $\beta$. An
exact solution for $r_{ta}$ is given in Appendix A of Eke et al.
(1996). For overdensities that are past turnaround the left-hand side
of equation (17) is integrated from zero to $r_{ta}$ and added to the
integral from $r$ to $r_{ta}$. The evolution of the radius of an
overdensity as a function of time is illustrated in Figure 1. Note
that the cosmological constant has the effect of slowing the
collapse as compared to a CDM universe.

\begin{figure}
\hbox{~}
\centerline{\psfig{file=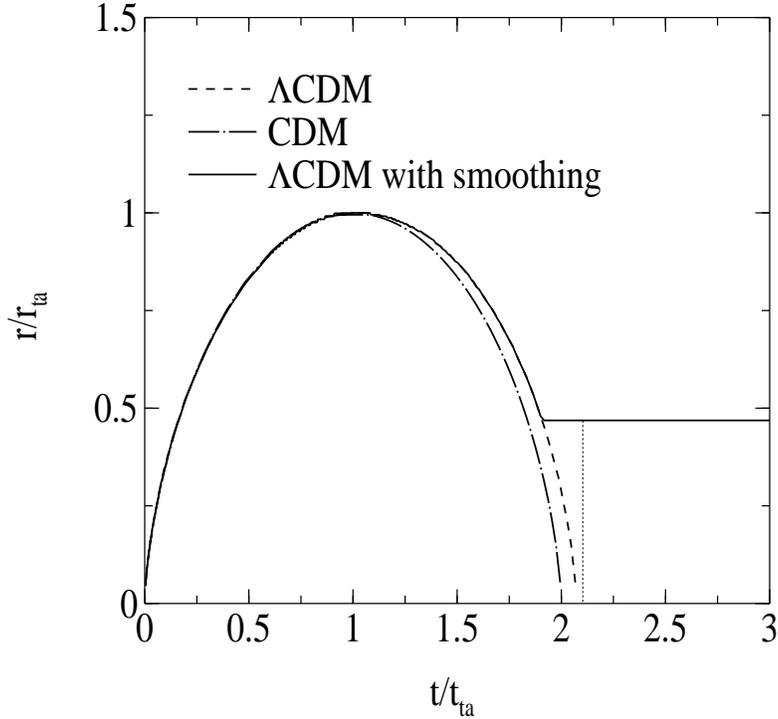,angle=0,width=5in}}
\vskip 1mm
\caption{The radial evolution of a density perturbation according to
  the STHC model. At the turnaround time, $t=t_{ta}$, the perturbation
  reaches a maximum-expansion radius and begins to collapse. As
  expected, in a $\Lambda$CDM cosmology (\emph{dashed} curve) the
  collapse takes somewhat longer than in a CDM cosmology
  (\emph{line-dot} curve). The collapse to a singularity
  predicted by the solution of the STHC model is avoided by the
  smoothing scheme (\emph{solid} curve) which yields a constant radius
  once the virialized overdensity is reached.}
\end{figure}

The non-linear overdensity is given by
\begin{equation}
1+\delta^{\rm{NL}} = \frac{\rho_{\rm{pert}}}{\rho_b},
\end{equation}
where $\rho_{\rm{pert}}$ is the mean density of the perturbed
region and $\rho_{b}$ is the background density at the given
redshift. Since $\rho_{\rm{pert}}=\rho_0 / r^3$ and $\rho_{b} =
\rho_0/ a^3$, the non-linear overdensity becomes
\begin{equation}
1+\delta^{\rm{NL}} = \left( \frac{a}{r} \right)^3.
\end{equation}
For a given non-linear overdensity of mass $M$ at redshift $z$ we can
find the radius of the perturbation $r$ such that $(S/N) \ge (S/N)_{\rm{min}}$.
We can then solve equation (17) for $\beta$.

We now relate this same $\beta$ to the linear-theory perturbation amplitude.
Eke et al.(1996) showed that
\begin{equation}
\beta = \frac{a_0(2\omega)^{1/3}}{3A(a_0(2\omega)^{1/3})}
\delta_0^{\rm{lin}},
\end{equation}
where $a_0$ is the scale factor today, $\delta_0^{\rm{lin}}$ is
the linear-theory overdensity extrapolated to the present and
\begin{equation}
A(x)=\frac{(x^3+2)^{1/2}}{x^{3/2}}\int^x
   \left( \frac{u}{u^3+2} \right)^{3/2} du,
\end{equation}
\cite{Pee80}. The linear-theory overdensity at redshift $z$ is given by
\begin{equation}
\delta^{\rm{lin}}(z) = \delta_0^{\rm{lin}} D(a),
\end{equation}
where $D(a)$, the linear theory growth factor for a $\Lambda$CDM
cosmology, is
\begin{equation}
D(a)=\frac{A(a(2\omega)^{1/3})}{A(a_0(2\omega)^{1/3})}.
\end{equation}
Using equations (20), (22), and (23) we get the desired relation
between the linear-theory overdensity and $\beta$:
\begin{equation}
\delta^{\rm{lin}}(z) = 3 \beta
\frac{A(a(2\omega)^{1/3})}{a_0(2\omega)^{1/3}}.
\end{equation}
Equations (17), (19) and (24) therefore provide a map between the
non-linear overdensity and the linear-theory overdensity at a given redshift.

It can be shown that $r \rightarrow 0$ in the limit that
$\delta^{\rm{lin}} \rightarrow \delta_c$,
corresponding to the well known infinite density predicted by the
solution of STHC. An actual overdensity will, of course,
virialize before reaching the singular solution. To properly
account for this we introduce the following smoothing scheme.

Rather than assuming that an overdensity satisfies equation (19)
throughout its evolution, assume it satisfies it only until it reaches
the virialized overdensity $1+ \delta_{\rm{vir}}^{\rm{NL}}(z)$. Once
the perturbation reaches the virialized overdensity take its radius to
be a constant with time so that the overdensity continues to grow only
because the cosmological background density keeps decreasing. The
non-linear overdensity is therefore given by
\begin{equation}
1+\delta^{\rm{NL}} = \left\{ \begin{array}{ll} \left( \frac{a(t)}{r}
\right)^3, & \mbox{if} \left( \frac{a(t)}{r} \right)^3 \leq
1+\delta_{\rm{vir}}^{\rm{NL}}(z);\\ (1+
\delta_{\rm{vir}}^{\rm{NL}})\left(    \frac{a(t)}{a_{\rm{vir}}}
\right)^3, & \mbox{otherwise}, \end{array}\right.
\end{equation}
where $a_{\rm{vir}}$ is the scale factor at virialization. Since
$\delta^{\rm{lin}}(t_2)=\delta^{\rm{lin}}(t_1) D(a_2)/D(a_1)$, the
linear-theory overdensity then becomes
\begin{equation}
\delta^{\rm{lin}} =
\left\{ \begin{array}{ll} 3 \beta
\frac{A(a(2\omega)^{1/3})}{a_0(2\omega)^{1/3}},  &  \mbox{if }
 1+\delta^{\rm{NL}}\leq 1+\delta_{\rm{vir}}^{\rm{NL}}(z);\\
\delta^{\rm{lin}}(a_{\rm{vir}}) \frac{D(a(t))}{D(a_{\rm{vir}})}, &
\mbox{otherwise}.\end{array} \right.
\end{equation}
Therefore, if the minimum non-linear overdensity needed to produce a
detectable weak-lensing signal at redshift $z$ is larger than the
virialization overdensity, we evaluate $a_{\rm{vir}}$ using equation
(25) and then compute the minimum linear-theory overdensity using the
lower expression in equation (26). In Figure 1 we plot the
radius of an overdensity as a function of time using this
smoothing scheme. In Figure 2 we show the non-linear overdensity
as a function of the linear-theory overdensity. Note that the value of
the overdensity at virialization can be obtained by assuming
$r=r_{\rm{vir}}$, the virialized radius, in equation (19), and using
the expression from Lahav et al. (1991) which gives the ratio between
the turnaround radius and the virialization radius. For convenience we
use the Kitayama \& Suto (1996) approximation to
$1+\delta_{\rm{vir}}^{\rm{NL}}(z)$, as well as their approximation to
$\delta_c(z)$. We independently verified that both approximations
matched the solution of the exact formalism described above.

\begin{figure}
\hbox{~}
\centerline{\psfig{file=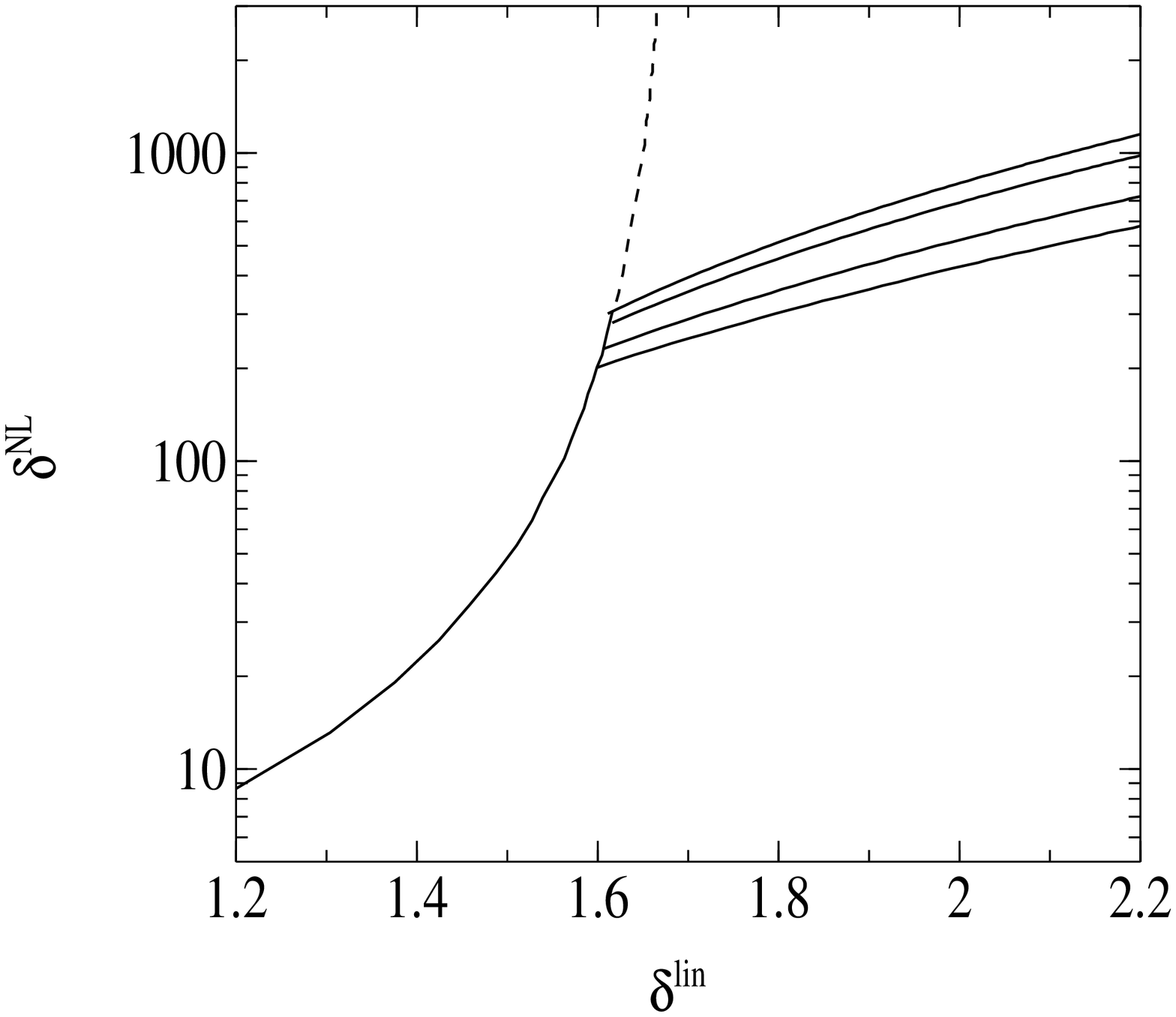,angle=0,width=5truein}}
\vskip 1mm
\caption{The nonlinear overdensity as a function of the linear-theory
overdensity according to the STHC model. The full solution of the STHC
model predicts collapse to an infinite overdensity as
$\delta_{\rm{lin}} \rightarrow 1.69$ (\emph{dashed} curve). According
to the smoothing scheme, however, once a mass concentration reaches
the virialization overdensity $1+\delta_{\rm{vir}}^{\rm{NL}}(z)$, its
radius remains constant so that the overdensity increases in
proportion to the decrease in the background density. The \emph{solid}
curves show the smoothing scheme solution for mass concentrations that
reach the virialization overdensity at $z=0.1, 0.2, 0.5$, and $1.0$,
from top to bottom. In an Einstein-de Sitter Universe, the virialization
overdensity is independent of redshift and therefore all of the solid
curves would be the same.}
\end{figure}

In summary, given the minimum non-linear overdensity needed to produce
a detectable weak-lensing signal,
$\delta^{\rm{NL}}_{\rm{min}}$, of an object of mass $M$ at
redshift $z$, we use equations (17), (25) and (26) to compute the
corresponding minimum linear-theory overdensity,
$\delta^{\rm{lin}}_{\rm{min}}$, needed to produce a detectable
weak-lensing signal. If $\delta^{\rm{lin}}_{\rm{min}} <
\delta_c(z)$ then the object can produce a
detectable weak gravitational lens, even though it is not yet
virialized.

\section{Abundances}

To calculate the abundance of overdensities that produce a
detectable weak-lensing signal as a function of redshift, we use
Press-Schechter theory assuming Gaussian statistics for the
initial linear-theory density field. The differential number
count of lensing objects per steradian, per unit redshift interval is
\begin{equation}
\frac{dN(\delta^{\rm{lin}}_{\rm{min}})}{dz d\Omega}=\frac{dN
(\delta^{\rm{lin}}_{\rm{min}})}{dV} \frac{dV}{dz d\Omega},
\end{equation}
where
\begin{equation}
\frac{dV}{dz d\Omega} = \frac{c}{H_0} \frac{(1+z)^2 D_{\rm{A}}(z)^2}
{\sqrt{\Omega_0(1+z)^3+1-\Omega_0}},
\end{equation}
is the comoving-volume element, $c$ is the speed of light,
$H_0$ is Hubble's constant, and $D_{\rm{A}}(z)$ is the angular-diameter
distance at redshift $z$. The total number density of weak lenses is 
given by
\begin{equation}
\frac{dN (\delta^{\rm{lin}}_{\rm{min}})}{dV} = \int_0^{\infty}
f(M; \delta^{\rm{lin}}_{\rm{min}}) \frac{dn}{dM}(M)dM,
\end{equation}
where $dn(M)/dM$, the comoving number density of virialized objects
of mass $M$ in the interval $dM$, is \cite{PreSch74}
\begin{equation}
\frac{dn}{dM}(M)=\sqrt{\frac{2}{\pi}} \frac{\rho_0}{M^2}
\frac{\delta_c(z)}{\sigma(M,z)} \left|\frac{d \ln\sigma}{d \ln
M}\right| \exp \left[-\frac{\delta_c(z)^2}{2 \sigma^2} \right].
\end{equation}
In this paper we use the Viana \& Liddle (1999) fits to the
dispersion of the density field, $\sigma (M,z)$, obtained from the
galaxy cluster X-ray temperature distribution function. The function
$f(M; \delta^{\rm{lin}}_{\rm{min}})$ is the fraction of objects,
either dark or virialized, that can lens ($\delta >
\delta^{\rm{lin}}_{\rm{min}}$) relative to those that are virialized ($\delta >
\delta_c$). The probability that an object's linear overdensity is in
the range $\delta_1 < \delta < \delta_2$ is
\begin{equation}
P(\delta_1 < \delta < \delta_2) =
\mbox{erf}\left(\frac{\delta_2}{\sqrt 2
\sigma(M,z)}\right) -
\mbox{erf}\left(\frac{\delta_1}{\sqrt 2 \sigma(M,z)}\right),
\end{equation}
where `erf' is the error function. Therefore, for
dark lenses (i.e., those objects with $\delta^{\rm{lin}}_{\rm{min}}<
\delta < \delta_c$)
\begin{eqnarray}
f_{\rm{dark}}(M,z) & = & \left\{
\begin{array}{ll} \frac{{\textstyle P(\delta^{\rm{lin}}_{\rm{min}} < \delta <
\delta_{\rm{c}})}} {{\textstyle P(\delta >\delta_{\rm{c}})}},       & 
\delta^{\rm{lin}}_{\rm{min}} < \delta_{\rm{c}};\\ \\
 0, &   \mbox{otherwise},\end{array} \right.
\end{eqnarray}
while for virialized lenses  ($\delta > \delta_c$ and $\delta
>\delta^{\rm{lin}}_{\rm{min}}$),
\begin{eqnarray}
f_{\rm{vir}}(M,z) & = & \left\{
\begin{array}{ll} \frac{\textstyle{ P(\delta > \delta^{\rm{lin}}_{\rm{min}})}}
 {{\textstyle P(\delta > \delta_{\rm{c}})}},     & \; \; \; \; \;\; \;
\delta^{\rm{lin}}_{\rm{min}} > \delta_{\rm{c}};\\ \\ 1, 
& \; \; \; \; \;\; \; \mbox{otherwise}.\end{array} \right.
\end{eqnarray}
For low enough masses, the minimum overdensity needed to produce a detectable
weak-lensing signal becomes so large that both $f_{\rm{dark}}$ and $f_{\rm{vir}}$
approach zero, thereby imposing an effective weak-lensing mass threshold 
(see Section 5.2). Integrating equation (27) over redshift assuming $f=f_{\rm{dark}}$ 
yields the number count of dark  lenses per unit area on the sky and similarly for
virialized lenses when $f=f_{\rm{vir}}$.

\section{Results}
\subsection{Minimum overdensity as a function of redshift}

\begin{figure}
\hbox{~}
\centerline{\psfig{file=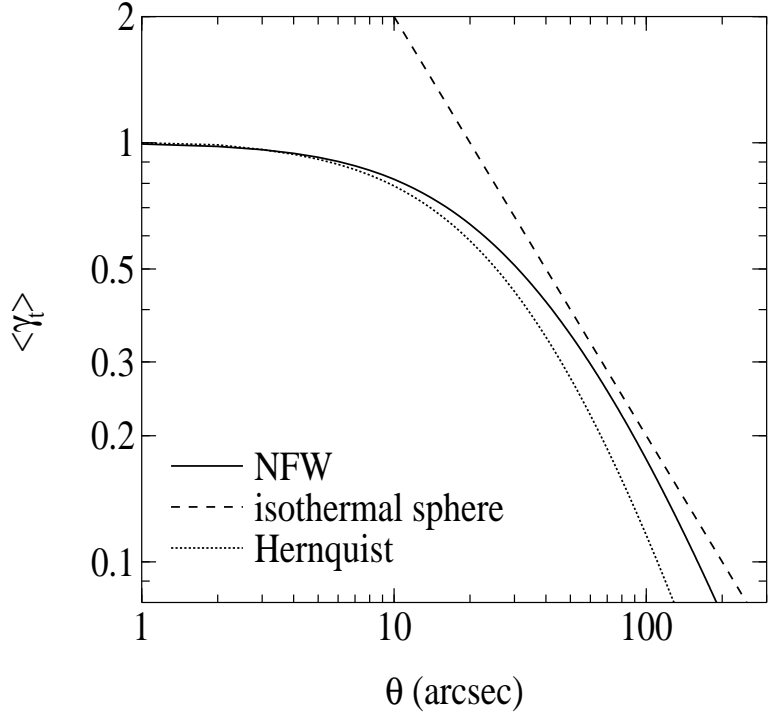,angle=0,width=5truein}}
\vskip 1mm
\caption{The mean tangential shear as a function of angular distance
from the lens center for an NFW (\emph{solid} curve), Hernquist
(\emph{dotted} curve) and isothermal sphere (\emph{dashed}
curve) density profile. The normalization is arbitrary.}
\end{figure}
We can now compute the sky density of weak lenses. To gain physical
insight into the results as well as illustrate the calculational
procedure discussed above we first show the redshift dependence of the
minimum non-linear overdensity. As noted earlier, the result is
sensitive to the lens density profile on account of the minimum
overdensity's dependence on the shear. Since the shear is proportional
to the surface mass density, the NFW and Hernquist profiles
(for whom $\rho \propto r^{-1}$ as $r \rightarrow 0$) have a
constant shear at small radii while the
isothermal sphere profile ($\rho \propto r^{-2}$) has a shear that
goes as $r^{-1}$ for all radii. This is shown in Figure 3, where we
plot the radial dependence of the mean tangential shear for these
different profiles.

In Figure 4, the minimum non-linear overdensity as a function of
redshift for a $10^{14} M_{\odot}$
object is plotted for the various profiles. All the profiles show
the same general trend: a minimum at $z \sim 0.3$ and monotonic rises
at lower and higher redshifts. This is a consequence of
the source-galaxy redshift distribution which we assume is given
by a function of the form
\begin{equation}
p_z(z_s)=\frac{\beta z_s^2}{\Gamma
(3/\beta)z_0^3}\exp\left[-(z_s/z_0)^{\beta}\right],
\end{equation}
with $\beta=1.5$ and mean redshift $\langle z_s \rangle \approx 1.5 z_0 = 1.2$
(cf. Smail et al. 1995; Brainerd et al. 1996; Cohen et al. 2000) . Since lenses are most
effective when they lie midway between the source
and the observer (i.e., the factor $D_d D_{ds} /D_s$ peaks when
$D_d \simeq D_{ds}$), an overdensity at $z \sim 0.3$ is ideally
positioned to lens source galaxies that are primarily located at
$z = \langle  z_s \rangle  \sim 1$, thereby accounting for the minimum
in the curves. Accordingly, overdensities located at lower and higher
redshifts than $z \sim 0.3$ are less effective at lensing so that a
larger overdensity is needed to produce a detectable lens. In addition,
for an overdensity with
redshift approaching unity, there are fewer background galaxies to
lens (less signal) as well as more foreground galaxies in the image
(greater noise), further decreasing the observed lensing
signal-to-noise ratio.

Another feature to note in Figure 4 is the difference in amplitude of
$1+\delta^{\rm{NL}}_{\rm{min}}$ between the different profiles.
Over most of the redshift range, the NFW profile requires the
smallest overdensity in order to produce a detectable
weak-lensing signal while the uniform-density sphere requires the
largest. This is because the NFW profile has its mass much more
centrally concentrated as compared to the uniform-density sphere.
A source galaxy at some angular radius near the lens center will
therefore be sheared more strongly by the former and hence
produce a larger signal. A similar explanation accounts for the
differences in amplitude of
$1+\delta^{\rm{NL}}_{\rm{min}}$ between the non-uniform profiles.

\begin{figure}
\hbox{~}
\centerline{\psfig{file=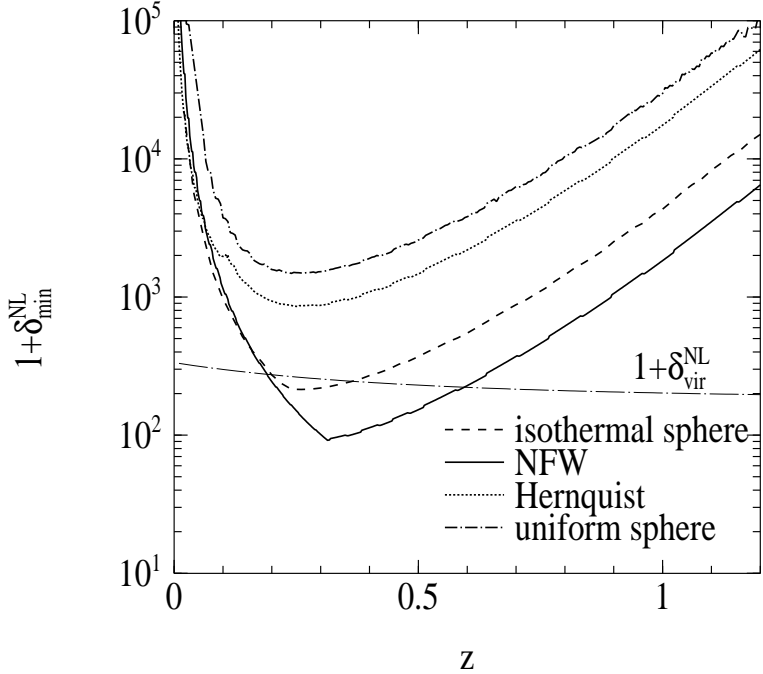,angle=0,width=5truein}}
\vskip 1mm
\caption{The minimum nonlinear overdensity needed to produce a
detectable weak lensing signal as a function of redshift for a
$10^{14} M_\odot$ object with a density profile that is a
uniform-density sphere (\emph{line-dot} curve), a truncated isothermal
sphere (\emph{dashed} curve), a Hernquist profile (\emph{dotted}
curve), and an NFW profile (\emph{solid} curve). An overdensity with
a larger mass will displace these curves downwards. The thin, \emph{long-dash-dot}
curve is the overdensity at virialization in the STHC model.}
\end{figure}

\subsection{The abundance of dark and virialized lenses}

In Figure 5 we show the redshift distribution
(normalized to unity) of dark and virialized lenses for the NFW,
Hernquist, and isothermal-sphere profiles.
\begin{figure}
\hbox{~}
\centerline{\psfig{file=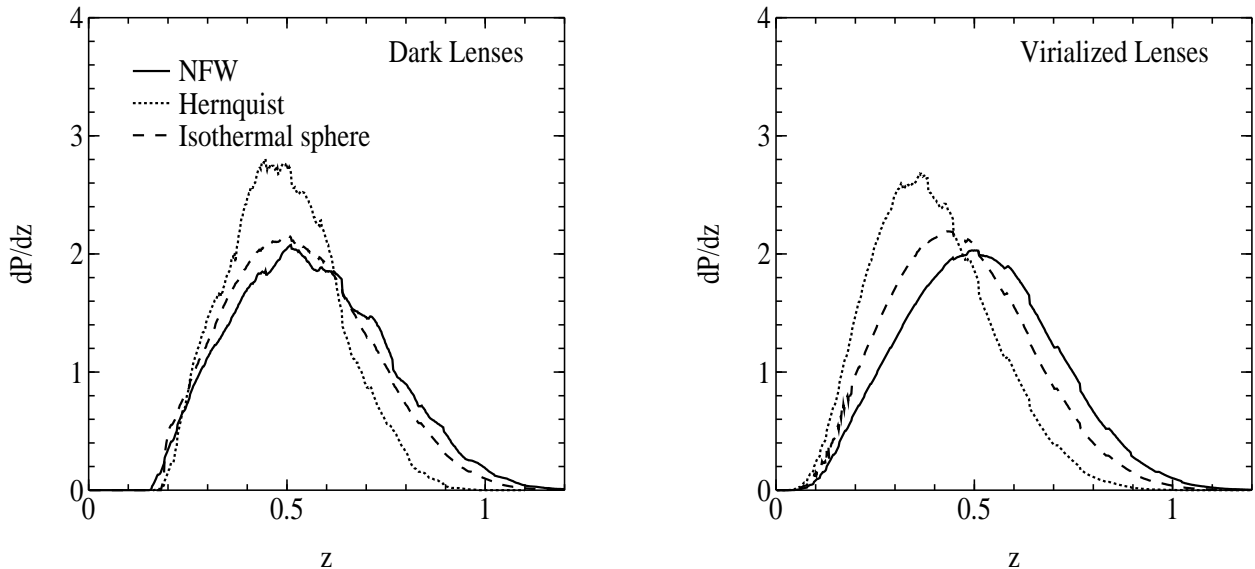,angle=270,width=7truein}}
\vskip 1mm
\caption{Redshift distribution of dark lenses (\emph{left} panel) and
virialized lenses (\emph{right} panel) for a truncated isothermal
sphere (\emph{dashed}  curve), an Hernquist profile (\emph{dotted}
curve), and an NFW  profile (\emph{solid} curve). The ordinate gives
the normalized probability distribution per unit redshift interval.}
\end{figure}
Because the minimum overdensity for the uniform-density
sphere was so large, the probability of detecting a lens with
such a profile is negligible and hence no longer considered. For all
three profiles the distribution peaks at $z \approx 0.5$ and has a
full-width at half-maximum of $\Delta z \approx 0.5$. The distribution
drops off at $z
\approx 1$ for two reasons: the minimum overdensity is becoming
increasing large since $\langle z_s \rangle \simeq 1$ and the STHC
dynamics predicts fewer and fewer massive, large overdensities at
these higher redshifts.
\begin{figure}
\hbox{~}
\centerline{\psfig{file=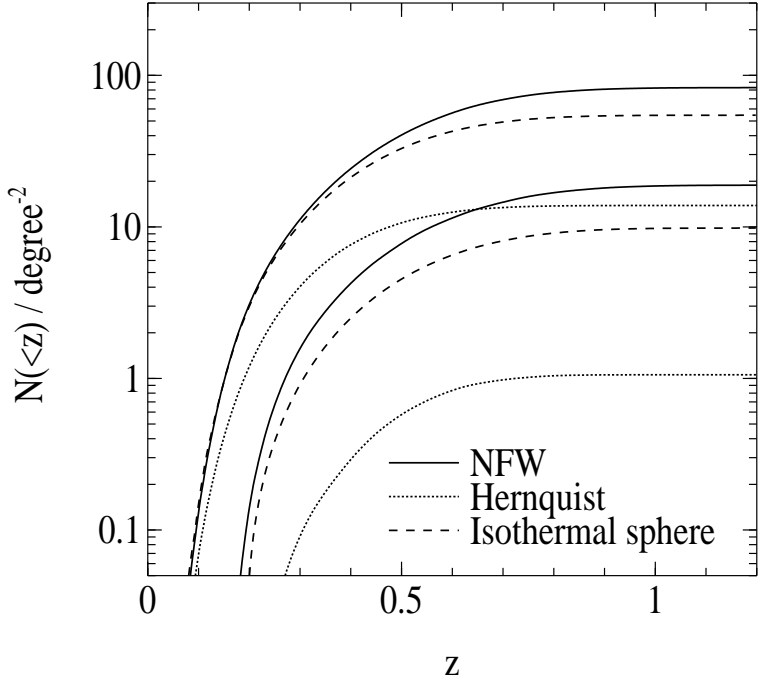,angle=0,width=5truein}}
\vskip 1mm
\caption{The number counts of dark and virialized lenses for a
truncated isothermal sphere (\emph{dashed} curve), an Hernquist
profile (\emph{dotted} curve), and an NFW profile (\emph{solid}
curve). The ordinate gives the sky density of lenses at redshifts less
than $z$. The top curve for a given density profile corresponds to the
sky density of virialized lenses and the bottom curve to the sky
density of dark lenses.}
\end{figure}

\begin{figure}
\hbox{~}
\centerline{\psfig{file=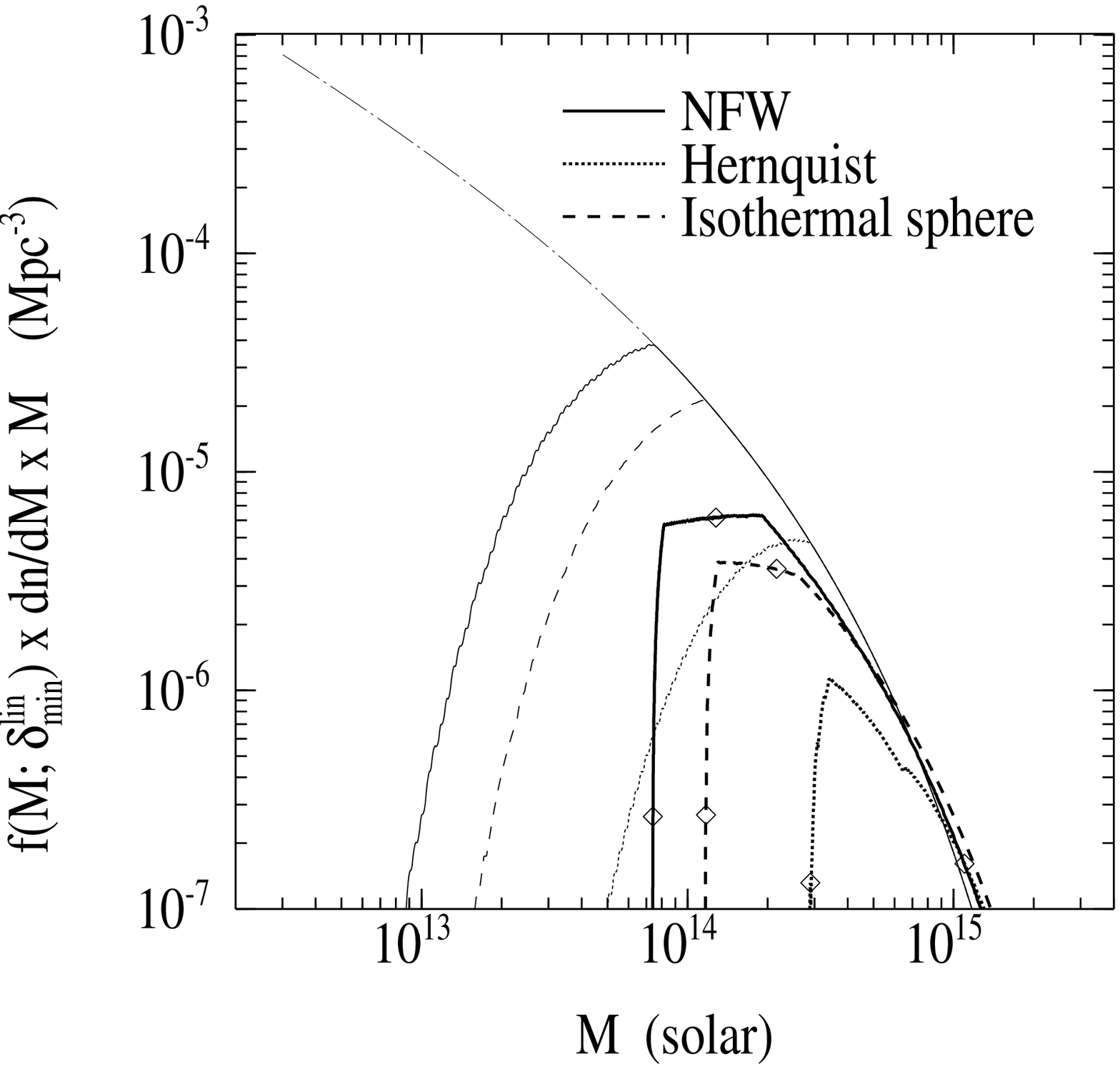,angle=0,width=5truein}}
\vskip 1mm
\caption{The predicted weak-lensing mass distribution at redshift $z=0.5$. 
Shown is the comoving number density as a function of mass of dark lenses 
(\emph{thick} lines) and virialized lenses
(\emph{thin} lines) of mass $M$ in the interval $d \ln M$.
Plotted for comparison is the virialized mass distribution,
i.e.,$\frac{dn}{dM}(M) \times M$ (\emph{thin}, \emph{dash-dot} line). Since the minimum
mass overdensity needed to produce a detectable lens (see Figure 4) is lowest for the NFW
profile (\emph{solid} curves), such a profile predicts a smaller weak-lensing mass 
threshold as compared to the Hernquist profile (\emph{dotted} curve)  and the truncated 
isothermal sphere (\emph{dashed} curve). The two diamonds on each dark lens mass distribution
curve mark the mass at which the minimum overdensity needed to produce a detectable lens is 
275 and 100. The sharp lower-mass cutoff in the dark lens mass distribution is a consequence of the
heaviside step-function nature of $f_{\rm{dark}}$.}
\end{figure}

The sky density of dark lenses as a function of redshift for the same
three profiles is shown in Figure 6. Depending on the density profile,
we expect to find between $1-20$ dark lenses per square degree out to
$z=1$ and virtually none at higher redshifts. The reason the Hernquist
profile predicts a smaller dark-lens sky density compared with the
isothermal sphere and NFW is that such a profile requires a larger
overdensity to produce a detectable weak-lensing signal (see Figure
3). Finally, note that although this distribution is integrated over
dark lenses of all masses, the minimum overdensity as a function of
redshift becomes so large for $M \la 5 \times 10^{13} M_{\odot}$
that there are virtually no dark lenses with such small masses.
This point is illustrated in Figure 7, where we plot the weak-lensing
mass distribution (i.e., the integrand of equation (29) times the mass) 
for both dark and virialized lenses at $z=0.5$. Furthermore,
since the Press-Schechter mass function falls off steeply
with mass, there will be very few dark lenses with $M \ga 10^{15} M_{\odot}$
despite the lower value of the minimum overdensity at these
masses.

In Figure 5 and 6 we also show the redshift distribution and sky
density of virialized lenses for the three different density profiles.
Although the weak-lensing mass threshold is, as expected, somewhat smaller
for virialized lenses than for dark lenses (see Figure 7) their normalized
distributions are
not very different. Nonetheless, the sky density of virialized lenses is $10-80$
degree$^{-2}$ and hence a factor of $4-10$ larger than the sky density
of dark lenses. This is because by redshifts of $z \approx 0.5$ (where
the distributions peak) a majority of objects in the mass range
that can lens will have already virialized.

\begin{figure}
\hbox{~}
\centerline{\psfig{file=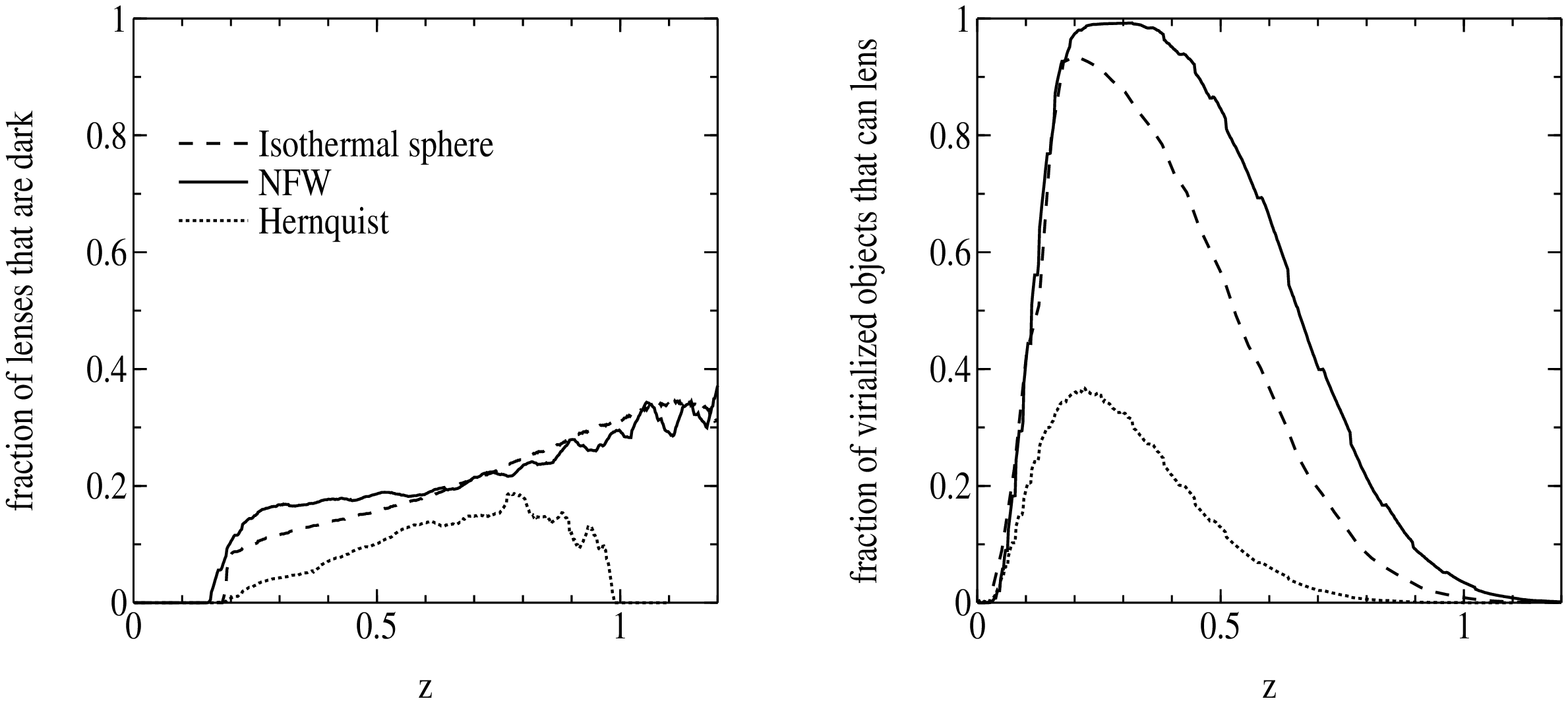,angle=270,width=7truein}}
\vskip 1mm
\caption{Left panel: The fraction of weak lenses that are dark lenses
as a function of redshift for a truncated isothermal sphere
(\emph{dashed} curve), an Hernquist profile (\emph{dotted} curve), and
an NFW profile (\emph{solid} curve). The fraction is relatively
constant between redshifts $z=0.2$ and $z=1.0$, beyond which
the abundance of both dark and virialized lenses drops to zero. The
coarseness of the curves for $z>1$ is an artifact of numerical noise that
is a result of this drop off in both abundances. Right panel: The fraction
of virialized objects with $M > 5 \times 10^{13} M_\odot$ that are
able to weak lens as a function of redshift for the same density
profiles as above.}
\end{figure}
Having computed the redshift distribution and sky density of dark and
virialized lenses we now determine their relative abundances. The
fraction of weak lenses that are caused by dark, non-virialized
objects as a function of redshift is shown in the left panel of Figure
8. Out to $z \approx 1$ the fraction is nearly constant with about
20\% of all weak lenses arising from dark objects. We again emphasize
that the predicted abundance of dark lenses relative to virialized lenses
is significant not because dark lenses comprise the lower-mass end of the
mass function; on the contrary, virialized lenses have a lower mass
threshold than dark lenses as shown in Figure 7. Rather, it is
significant because according to
the STHC model, a substantial fraction of cluster-mass objects are sufficiently
overdense to produce a detectable lensing signal despite not having
reached the virialization overdensity. For $z > 1$ the
abundance of weak lenses of all types (both virialized and dark) drops
off significantly. This, again, is because $\langle z_s \rangle \sim 1$
and because the evolution of overdensities has not yet had enough time
to produce sufficiently large overdensities. This is also illustrated in the
right panel of Figure 8 where we show the fraction of virialized
objects that can lens as a function of redshift. For $0.2 \la z \la 0.5$
a large fraction of virialized objects with $M > 5 \times 10^{13} M_{\odot}$
can produce a detectable weak-lensing signal but, for the same reason as
above, by $z=1$ this fraction is nearly zero.

Finally we would like to point out that given the above results for the
weak-lensing mass distribution, it is not surprising that in their study of 
weak lensing by low-mass galaxy groups, Hoekstra et al. (2001) could only 
(just barely) detect a weak-lensing signal by stacking 50 such groups together. 
Namely, the groups in Hoekstra et al.'s sample,
which were at a mean redshift of 0.3, had a mean overdensity of only
$\sim$ 75 and a mean mass of just $\sim$
$4 \times 10^{13} M_{\odot}$, assuming an isothermal density profile and using
their measured value of $\sim$ 275 km s$^{-1}$ for the lensing-inferred
velocity dispersion. Therefore, as Figure 7 suggests, individual groups from
their sample were neither massive enough nor sufficiently overdense to produce
a detectable weak-lensing signal.

\subsection{The effect of increasing the image size on the lensing
signal}

In the above calculations we assume that the lensing images are 5
arcmin in radius, roughly the size of lensing maps to date. However,
if a lens is relatively nearby or has a large radial extent it is
possible that a large fraction of the total lensing signal is
missed. This effect might be especially troublesome for the
detection of dark lenses, given that they are not yet
virialized and hence have larger radii. We now address this
issue by determining the extent to which increasing the image size
alters the predicted abundance of dark lenses. 

Before moving on however, we note that while we examined the predicted
distribution and sky density of weak lenses for a variety of profiles, there
is good reason to regard the NFW profile as the most plausible. For virialized
lenses this is clearly the case as N-body simulations
show that the halo density profiles are well fit by the NFW form. Though it is
difficult to be as certain in the case of dark, non-virialized, lenses 
(N-body simulation fits to profiles have so far only been for virialized systems), 
because most of the dark
lenses are well past turnaround ($1+\delta^{\rm{NL}} \ga 50$) and because it is
unlikely that the STHC model perfectly describes the evolution of overdensities
all the way to virialization, assuming an NFW profile for dark lenses is a
fair approximation. Furthermore, since virialization is expected to occur from
inside-out, the centers of dark lenses, where most of the lensing signal is coming
from (as we show quantitatively below), are likely near virialization and hence well
described by the NFW profile. For these reasons (and also to avoid overly cluttered
figures), the rest of the figures in this paper show results only for the NFW profile.
To obtain approximate results for the other profiles, simply scale by the relative
abundances shown in Figure 6.

\begin{figure}
\hbox{~}
\centerline{\psfig{file=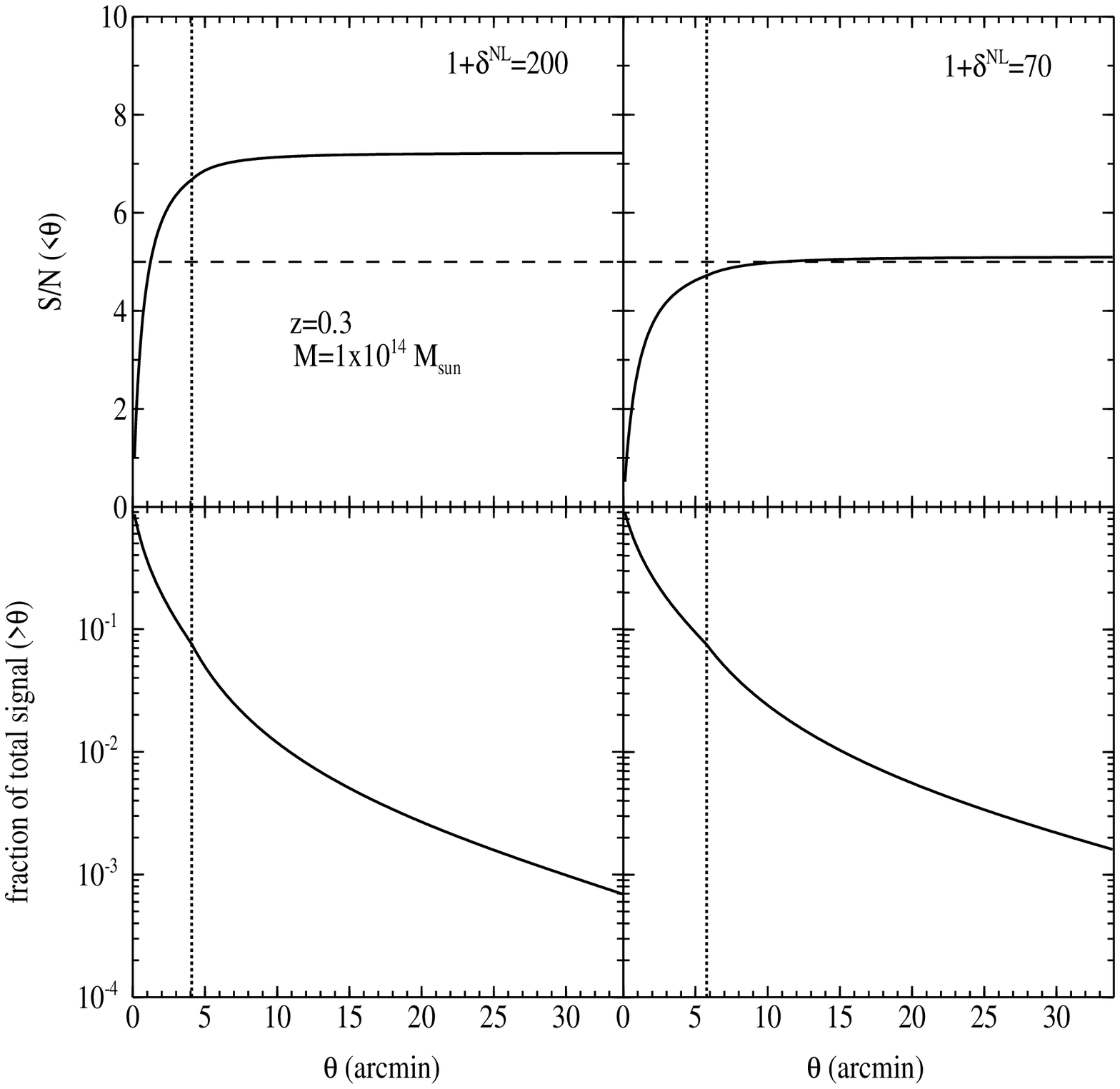,angle=0,width=5truein}}
\vskip 3mm
\caption{Upper panels: The signal-to-noise ratio within an angular
radius $\theta$ from the lens
center as a function of $\theta$ for an NFW profile. Lower panels: The
fraction of the total lensing signal that comes from outside the
angular radius $\theta$. Approximately 90\% of the lensing signal
comes from a region smaller than the lensing halo radius. All four
panels correspond to an object at redshift $z=0.3$ and mass
$M=10^{14} M_\odot$. The left-hand-side plots are for a nonlinear
overdensity of 200 and the right-hand-side plots for a nonlinear
overdensity of 70.}
\end{figure}

\begin{figure}
\hbox{\hskip 0.2truein \vbox{\vskip 0.0truein
\psfig{file=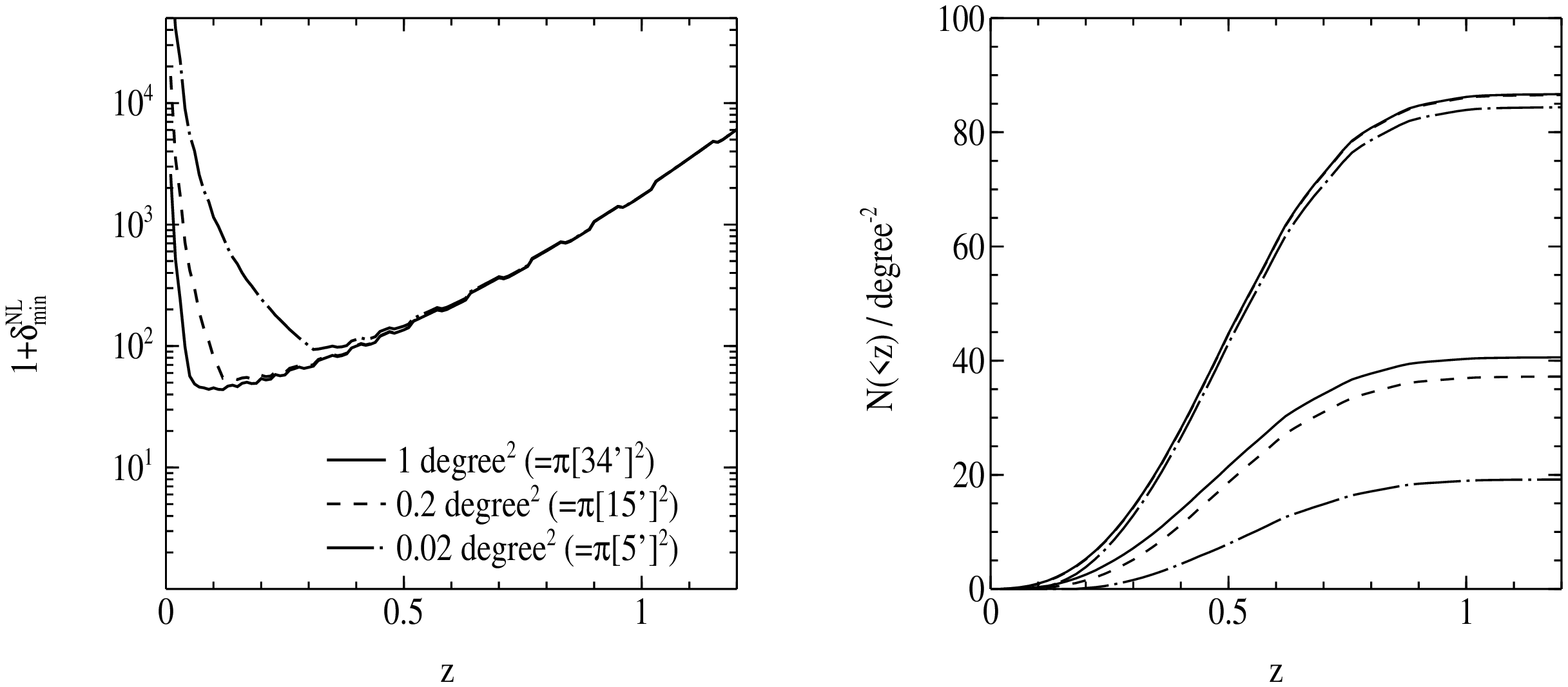,angle=270,width=7truein}}}
\caption{The effect of increasing the image size on the
abundance estimates. Plotted are the minimum overdensity needed to
produce a detectable lens (\emph{left} panel) and the sky density of
weak lenses (\emph{right} panel) as functions of redshift for image
sizes of 0.02 degree$^2$ (\emph{line-dot} curve), 0.2 degree$^2$
(\emph{dashed} curve), and 1 degree$^2$ (\emph{solid} curve). The top
three curves in the right panel correspond to virialized lenses and
the bottom three to dark lenses. An NFW profile is assumed. Note that
increasing the image size beyond 1 degree$^2$ barely increases the
predicted sky density since the signal becomes increasingly small at
larger angular distances from the lens center (see Figure 9).}
\end{figure}

In Figure 9 we plot the cumulative signal-to-noise ratio and the
fraction of the total signal as a function of the angular distance
from the lens center for a lens with an NFW profile at redshift
$z=0.3$ with mass $M=10^{14} M_\odot$. Although the fraction of the
signal that comes from within 5 arcmin is $\sim 90$\% for lenses of
overdensity $1+\delta^{\rm{NL}}_{\rm{min}}=200$ and
$1+\delta^{\rm{NL}}_{\rm{min}}=70$ the lens with overdensity
200 requires an image size of just $\sim$ 2 arcmin to be detectable
($S/N=5$) while the lens with overdensity 70 requires $\sim 10$ arcmin
to be detectable. In general we find that in order to detect nearly
all dark lenses with  $S/N \ge 5$ in a given field the image area must
be  at least $\sim \pi (15')^2 \approx 0.2$ degree$^2$, as shown
in Figure 10. Larger image sizes will not significantly increase the
number of dark lenses detected as very little signal comes from radii
larger than 15 arcmin. Also note that although lensing geometry favors
a lens midway between observer and source, this effect is somewhat
countered by the fact that the closer a weak lens is to the observer,
the closer the source galaxy images pass to the highly overdense lens
center (i.e., the solid angle subtended by the lens is larger). As a
result, if the image size is large enough to enclose a large portion
of the lens core, the lensing signal will be most strong when the
lens-observer distance is slightly smaller than the lens-source
distance. This accounts for the shift, shown in Figure 10,  of the
minimum of $1 + \delta^{\rm{NL}}_{\rm{min}}(z)$ toward smaller
redshift as the image size is increased.

\subsection{Estimating $\sigma_8$ from the abundance of weak lenses}

\begin{figure}
\hbox{~}
\centerline{\psfig{file=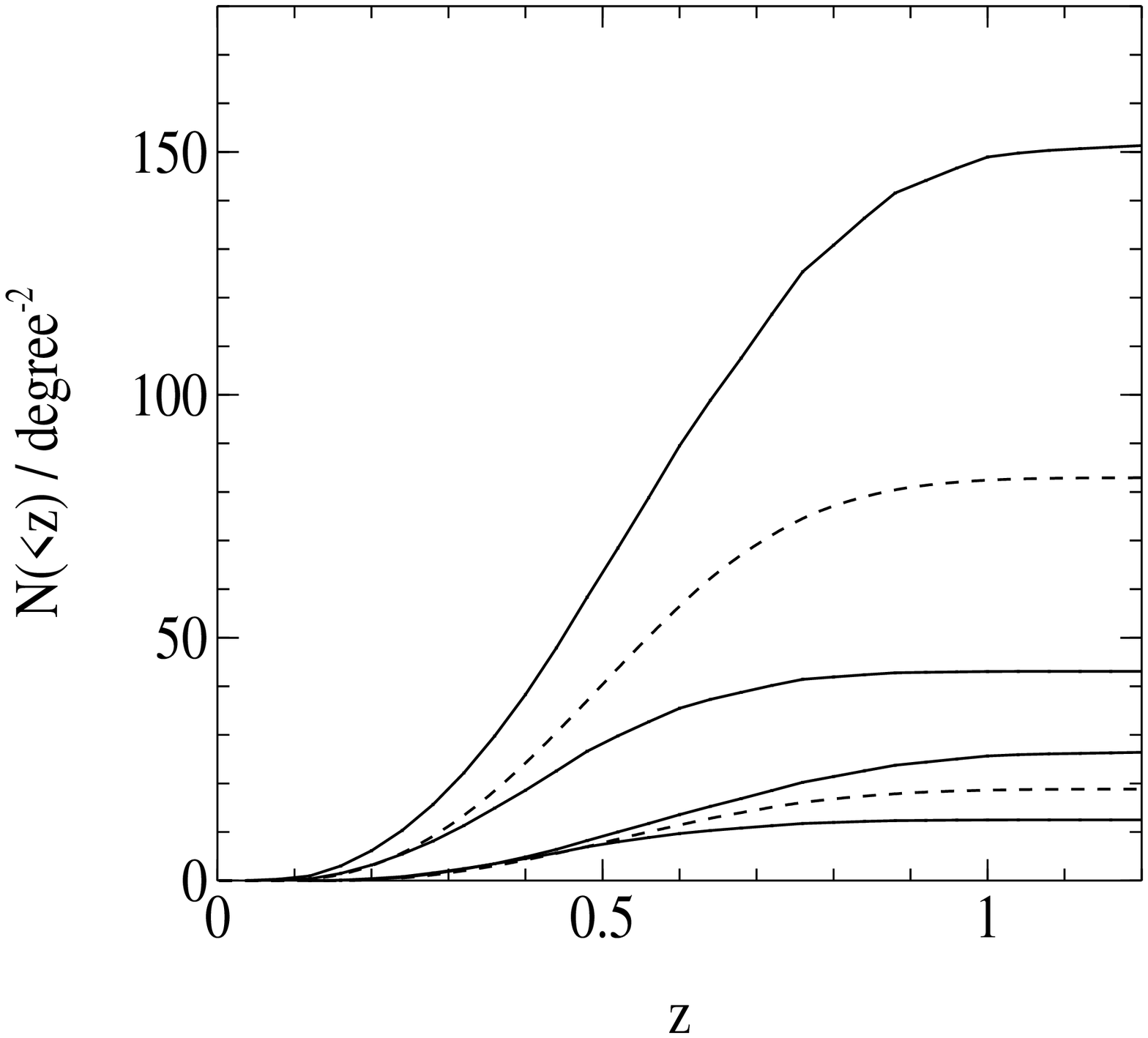,angle=0,width=5truein}}
\vskip 1mm
\caption{The predicted sky density of weak lenses at the 95\%
confidence limits of $\sigma_8$ (\emph{solid} curves) given by Viana
\& Liddle (1999). The \emph{dashed} curves are the predicted sky
densities for the mean value of $\sigma_8$. The top three curves
correspond to virialized lenses and the bottom three to dark lenses.
An NFW profile is assumed.}
\end{figure}
The present-day abundance of rich, X-ray clusters has been used to constrain
the value of $\sigma_8$, the amplitude of mass fluctuations in spheres
of radius 8$h^{-1}$ Mpc (Evrard 1989; Henry \& Arnaud 1991; White,
Efstathiou \& Frenk 1993; Viana \& Liddle 1996; Eke, Cole \& Frenk
1996; Kitayama \& Suto 1997, Viana \& Liddle 1999). In Figure 11 we
show the extent to which the measured abundance of weak lenses (from
future cosmic-shear surveys, say) can further constrain $\sigma_8$.
Here we have plotted the sky density of weak lenses (both virialized
and dark) at the 95\% confidence limits of $\sigma_8$ given by Viana
\& Liddle (1999). Since weak lenses are produced by
only relatively rare objects, their abundance is very sensitive to the
value of $\sigma_8$, suggesting the usefulness
of weak lenses in measuring the amplitude of mass fluctuations.

Another benefit of using weak lenses to measure $\sigma_8$ is their broad redshift
distribution. In particular, a systematic uncertainty in measuring $\sigma_8$ by measuring
rich cluster abundances is the degeneracy between $\sigma_8$ and $\Omega_m$
that arises from the limited range in redshift in which rich clusters are
observed. To break this degeneracy substantial effort is made to measure not
only the present-day rich cluster abundance but also the rich cluster abundance at
higher redshifts ($z \approx 0.3$; e.g., Henry 1997). This, in turn, gives an
estimate of the evolution of the cluster mass function and hence an estimate
of $\Omega_m$. However, because they are faint, detecting high redshift
($z > 0.3$) rich clusters is difficult. Weak lenses, on the other hand,
do not suffer from this limitation and in fact are expected to have a broad
redshift distribution and be most abundant at $z \sim 0.5$ (see Figure 5).
As a result, detecting weak lenses provides an excellent means of measuring the
evolution of the mass function and hence measuring $\Omega_m$. By thereby breaking
the degeneracy between $\sigma_8$ and $\Omega_m$, weak lensing surveys are
also well-suited to constrain the power-spectrum amplitude, $\sigma_8$.

\section{Discussion and Conclusions}

In this paper, we have calculated the abundance of dark and
virialized lenses. This was accomplished by first expressing the
lensing signal strength as a function of the dark-matter overdensity
and redshift. Having determined the overdensity required to produce a
detectable weak-lensing signal we used the STHC model to calculate the
differential abundance of overdensities as a function of position
along their evolutionary cycle. Overdensities whose lensing
signal yielded $S/N \ge 5$ were divided into two classes: those with
$1+\delta^{\rm{NL}}<1+\delta^{\rm{NL}}_{\rm{vir}}$ were dark lenses
while those with $1+\delta^{\rm{NL}}>1+\delta^{\rm{NL}}_{\rm{vir}}$
were virialized lenses.

The distinction between dark and virialized lenses was based on the former being
at an unrelaxed, and hence earlier, stage in the overdensity
evolutionary cycle. This distinction is not arbitrary but rather is expected to result in
observational features that definitively separate the two classes of lenses. For instance,
since dark lenses will typically have overdensities of $1+\delta^{\rm{NL}} \sim 100$
while virialized lenses have $1+\delta^{\rm{NL}} \sim 300$ (see Figure 4), the
projected surface density of a dark lens is smaller than that in a virialized lens by a
factor of $3^{2/3} \sim 2$. The sky density of galaxies in a dark lens will therefore be
about two times smaller than in a virialized lens. As it is difficult to detect a significant
galaxy overdensity for even a virialized, lensing, cluster at redshifts of $z \sim 0.5$, it
will be all the more difficult to do so for a dark cluster. Another distinctive observational
feature expected of dark lenses is a low X-ray luminosity as compared
with virialized lenses, a consequence of the X-ray luminosity
function's steep dependence on total virialized mass. This effect
might also account for the low X-ray luminosities observed by Postman
et al. (2001) in three high-redshift clusters; namely, these objects
are in fact proto-clusters that have not yet completely virialized.

Although we considered a variety of density profiles in our calculations of the
predicted distribution and sky density of weak lenses, as we noted in Section 5.3,
there is good reason to regard the NFW profile as the most plausible form for both
virialized and dark lenses. Nonetheless, while N-body simulations
show that virialized systems are well fit by the NFW form, testing whether non-virialized,
cluster-mass, halos in N-body simulations are also well described by the NFW profile 
is a worthwhile investigation that has not yet been performed. That said, 
we have shown that the redshift distribution of dark and virialized lenses for all 
the considered profiles is fairly broad with an average around $z=0.5$ and a
FWHM of $\Delta z \approx 0.5$. The sky density of dark lenses for the NFW profile
was calculated to be $\sim 20$ degree$^{-2}$ (and $\sim 10$ degree$^{-2}$ for an
isothermal sphere profile) and should therefore be readily detectable by
upcoming cosmic shear surveys. For virialized lenses, we found a sky density of
$\sim 80$ degree$^{-2}$ assuming an NFW profile (and $\sim 50$ degree$^{-2}$ 
for an isothermal sphere profile), a factor of 4 to 5 larger than that of dark
lenses. This difference is due to the fact that most of the weak lenses are at 
redshift $z \approx 0.5$ and have masses of $\sim 10^{14} M_{\odot}$ so that the 
majority are, according to the STHC model, virialized. It is important to note 
that while the aperture mass weight function used here was chosen to match a specific 
density profile, as we show in an 
upcoming paper (Weinberg \& Kamionkowski 2002), a more general, non-optimal weight function, 
such as that given by Schneider et al. (1998), would lower the overall abundance 
of both types of lenses equally. The principle result of this paper, namely the expectation
that $\sim 10-20$\%  of weak-lenses are dark, would not change. 

We find it encouraging that given the sky coverage of weak-lensing maps to date 
($\sim 1000$ arcmin$^2$) and the average size of the individual lensing maps ($\sim 30$ 
arcmin$^2$ ), the number of dark lenses we would expect to have seen is of order unity 
and thus consistent with the detection (Erben et al. 2000, Miralles et al. 2002) of one or 
two dark lenses. Furthermore, in mock observations of numerical simulations, 
White, van Waerbeke, \& Mackey (2001) showed that a weak-lensing search for clusters 
will likely suffer from serious line-of-sight projection effects due to the fact that 
clusters preferentially live in larger structures. These structures on larger scales, 
which perhaps correspond to 2$\sigma$ -- 3$\sigma$ peaks in the primordial distribution, 
may well be the type of systems that we find give rise to dark lenses.

Finally, we have also shown that measuring the abundance of weak lenses can
substantially help to constrain $\sigma_8$, the rms mass fluctuation
in spheres of radius 8$h^{-1}$ Mpc. This is a consequence of the broad
redshift distribution of weak lenses and the fact that they correspond to
high-density peaks in the Gaussian primordial distribution. Cosmic-shear surveys, with
their ability to detect cluster mass weak lenses over large areas
of sky, should therefore provide a powerful new technique for
determining the power-spectrum amplitude.

\section*{ACKNOWLEDGMENTS}

We thank R. Ellis for useful comments and an anonymous referee for very
constructive suggestions and comments which have improved the presentation 
of this paper. NNW acknowledges the support of an NSF Graduate Fellowship.  
This work was supported in part by NSF AST-0096023, NASA NAG5-8506, and DoE
DE-FG03-92-ER40701.

\appendix

\section{Derivation of the signal-to-noise relation for various
density profiles}

Starting from equation (14), we derive the signal-to-noise
relation for a point mass, a uniform-density  sphere, a truncated
isothermal sphere, an NFW profile and a Hernquist profile. Since these
profiles are all axially-symmetric, $\langle \kappa \rangle (\theta) =
\kappa (\theta)$.

\begin{enumerate}
\item \emph{Point Mass}: The dimensionless mean surface mass density within
a circle of radius $\theta$ for a deflecting lens of point mass $M$ at
angular diameter distance $D_d$ is
\begin{equation}
\bar{\kappa} = \frac{1}{\Sigma_{\rm{crit}_{\infty}}} \frac{M}{\pi P^2},
\end{equation}
where $P=\theta D_d$. The quantity $\kappa$, the dimensionless mean surface mass
density on a circle of radius $\theta$, is $\propto \delta(\theta)$,
the Dirac delta function.  Therefore, by equation (14), the signal-to-noise
relation for a point mass is given by
\begin{equation}
\frac{S}{N} = \frac{\sqrt 2 \langle Z \rangle M}{\sigma_{\epsilon}
        \Sigma_{\rm{crit}_{\infty}}\pi D_d^2 }
  \frac{\sqrt{\pi n}}{\theta_{\rm{out}}}
  \sqrt{\left(\frac{\theta_{\rm{out}}}{\theta_{\rm{in}}}\right)^2-1}.
\end{equation}
This can be expressed as a minimum mass needed to produce a detectable
weak-lensing signal, which in useful units is
\begin{eqnarray*}
M_{\rm{min}}=3.7 \times 10^{13}
\left(\frac{S/N}{5}\right) \left(\frac{D_d}{0.3 D_H}\right)^2
\left(\frac{\sigma_{\epsilon}}{0.2}\right)
\left(\frac{\theta_{\rm{out}}}{5 \; \mbox{arcmin}}\right) \left(\frac{n}{30 \;
\mbox{arcmin$^{-2}$}}\right)^{-1/2}
\\ \times
\left(\frac{(\theta_{\rm{out}}/\theta_{\rm{in}})^2-1}{100}\right)^{-1/
2} \Sigma_{\rm{crit}_{\infty}} \langle Z \rangle^{-1} M_\odot,
\end{eqnarray*}
where $D_H=c/H_0$ is the Hubble distance.

\item \emph{Uniform Density Sphere}: Repeating the same procedure as
above but for a sphere of uniform density $\rho(r) = \rho_c$ and mass
$M$ yields the following for the surface mass density (where we use
the Abel integral equation to relate volume mass density to surface
mass density);
\begin{eqnarray}
\kappa & = & \frac{1}{\Sigma_{\rm{crit}_{\infty}}}
\int_{-\infty}^{\infty} dz \; \rho(r) = \frac{2
\rho_c}{\Sigma_{\rm{crit}_{\infty}}} \sqrt{R^2 - P^2}, \\ \bar{\kappa}
& = & \frac{1}{\pi P^2} \int_0^P \kappa(P') \, 2 \pi P' \, dP'
\nonumber \\ &=& \frac{4 \rho_c}{3 \Sigma_{\rm{crit}_{\infty}}} \left(
\frac{R^3 - -(R^2-P^2)^{3/2}}{P^2} \right),
\end{eqnarray}
where $dz$ is along the line of sight and $R=(3M/4\pi \rho_c)^{1/3}$
is the radius of the sphere. The signal-to-noise ratio is then
computed by solving equation (14) with the above relations for
$\kappa$ and $\bar{\kappa}$.

\item \emph{Truncated Isothermal Sphere}: The radial density profile of an
isothermal sphere is
\begin{equation}
\rho(r)=\frac{\sigma_v^2}{2 \pi G r^2},
\end{equation}
where $\sigma_v$ is the line-of-sight velocity dispersion of the
particles (i.e., galaxies) in the system. The surface mass
density is then given by
\begin{equation}
\kappa = \frac{1}{2} \bar{\kappa} = \frac{\theta_E}{2 \theta},
\end{equation}
where $\theta_E= \sigma_v^2 / G D_d \Sigma_{\rm{crit}_{\infty}}$.
For a truncated isothermal sphere of mass $M$ and radius
$R$, $M=\int_0^R dr \rho(r)4\pi r^2 = 2 \sigma_v^2 R/G$, so that
\begin{equation}
\theta_E= \frac{M}{2 R} \frac{1}{ D_d \Sigma_{\rm{crit}_{\infty}}}.
\nonumber
\end{equation}
Equation (14) then gives
\begin{equation}
\frac{S}{N} = \frac{\langle Z \rangle M}{\sigma_{\epsilon}
\Sigma_{\rm{crit}_{\infty}} D_d} \frac{\sqrt{\pi n}}{2 R}
\sqrt{\ln(\theta_{\rm{out}}/ \theta_{\rm{in}}}).
\end{equation}

\item \emph{NFW Profile}: The NFW density profile is given by
\begin{equation}
\rho(r)=\frac{\rho_s}{(r/r_s)(1+r/r_s)^2},
\end{equation}
where $r_s$ and $\rho_s$ are the scale radius and density,
respectively. The mass within radius $r$ is then
\begin{equation}
M(r)= 4 \pi \rho_s r_s^3 \left(
\ln(1+r/r_s) - \frac{r/r_s}{1+r/r_s} \right).
\end{equation}
Bartelmann (1996) (see also Wright \& Brainerd 2000) showed that the
radial dependence of the tangential shear for an NFW profile is
\begin{equation}
\gamma_{\rm{nfw}}(x)=\bar{\kappa}(x)-\kappa(x)=
\frac{\rho_s r_s}{\Sigma_{\rm{crit}_{\infty}}} g(x),
\end{equation}
where $x=\theta D_d / r_s$ and
\begin{equation}
g(x)= \left\{ \begin{array}{ll}     
\frac{8{\rm arctanh}
\sqrt{\frac{1-x}{1+x}}}{x^{2}\sqrt{1-x^{2}}}
+ \frac{4}{x^{2}}\ln\left(\frac{x}{2}\right) - \frac{2}{\left(x^{2}-1\right)}+
\frac{4{\rm arctanh}\sqrt{\frac{1-x}{1+x}}}{\left(x^{2}-1\right)\left(1-x^{2}\right)^{1/2}}
,\hspace{1.0cm}  \mbox{$\left(x < 1\right)$} \\
 & \\ \frac{8\arctan\sqrt{\frac{x-1}{1+x}}}{x^{2}\sqrt{x^{2}-1}}
\hspace{0.1cm}
+ \hspace{0.1cm}\frac{4}{x^{2}}\ln\left(\frac{x}{2}\right) - \frac{2}{\left(x^{2}-1\right)}
\hspace{0.1cm}+\hspace{0.1cm}
\frac{4\arctan\sqrt{\frac{x-1}{1+x}}}{\left(x^{2}-1\right)^{3/2}}
\hspace{1.0cm} \mbox{$\left(x > 1\right)$}.
\end{array}
\right.
\end{equation}
The signal-to-noise ratio is then
\begin{equation}
\frac{S}{N}=\frac{2 \sqrt{\pi n} \langle Z
\rangle}{\sigma_{\epsilon} \Sigma_{\rm{crit}_{\infty}} D_d} \rho_s r_s^2
\sqrt{\int_{x_{\rm {in}}}^{x_{\rm {out}}}dx \, x \, g(x)^2}.
\end{equation}

There are thus three unknowns if given an overdensity of mass $M$:
$r_s$, $\rho_s$ and $R$. We therefore need a third relation in
addition to Equations (A10) and (A13) in order to break the
degeneracy. It is obtained via the following conservation of energy
argument, first put forth by Dalcanton et al. (1997) for the case of
disk formation.

Assume the mass profile before collapse is a uniform sphere of radius
$R_i$ and assume that at this initial stage the system's energy is
entirely gravitational ($E = -3GM^2/5R_i$). As noted by Dalcanton et al.
(1997), this assumption is well motivated in the context of disk
formation by the observed similarity between disk angular momentum
distributions and the angular momentum distribution of a uniformly
rotating sphere. It is natural to assume a similar initial condition
occurs for systems at larger scales, i.e., cluster masses. As the
overdensity collapses and approaches virialization, the mass
distribution evolves into an NFW profile, as suggested by numerical
simulations. At this stage the systems potential energy within a
radius $r$ is
\begin{equation}
\Phi(y) = -8 \pi^2 G \rho_s^2 r_s^5
\left(1 - \frac{2 y \ln y +1}{y^2}\right),
\end{equation}
where $y \equiv 1+ r/r_s$. Assuming the energy of the
overdensity within $R_i$ is conserved during collapse and that the
system is near virialization so that $E \approx \left| \Phi
\right|/2$ then gives $R_i=8.74 r_s$. Since the truncation radius is
given by the radius that contains mass $M$, by conservation
of mass $R=R_i=8.74 r_s$ (i.e., though the mass is redistributed as
the overdensity evolves the size of the sphere containing mass $M$ is
constant in time). When we include the effects of the cosmological
constant in the conservation of energy argument there is little
change in the result. The above relation between $R$ and $r_s$ thus
provides the sought after third equation needed to break the
degeneracy between $r_s$, $\rho_s$ and $R$. In an upcoming paper
(Weinberg \& Kamionkowski 2002) we show that the above approach yields
concentration parameters that are slightly different from those
obtained by N-body simulations (i.e., Bullock et al. 2001).
Nonetheless, the concentration parameters obtained by the two
approaches predict a similar abundance of virialized lenses. Note that
since the N-body simulations fit the concentration parameters to
virialized objects, the above analytic approach must be used in order
to compute the abundances of dark lenses.

\item \emph{Hernquist Profile}: The Hernquist profile is given by
\begin{equation}
\rho(r)=\frac{M_{\infty}}{2 \pi}\frac{1}{(r/r_s)(r+r_s)^3},
\end{equation}
where $r_s$ is the scale radius and $M_{\infty}$ is the mass
enclosed at infinity. The mass within radius $r$ is then
\begin{equation}
M(r)=M_{\infty}\left(\frac{r/r_s}{1+r/r_s}\right)^2.
\end{equation}
Using the Abel integral equation it can be shown that the
dimensionless surface mass density for the Hernquist profile is
\begin{equation}
\kappa(x)=\frac{M_{\infty}}{\pi r_s^2 \Sigma_{\rm{crit}_{\infty}}} f(x),
\end{equation}
where $x=\theta D_d / r_s$ and
\begin{equation}
f(x)= \left\{ \begin{array}{ll}
\frac{1}{(x^2-1)^2}\left[\frac{(2+x^2){\rm
arctanh}\sqrt{\frac{1-x}{1+x}}}
{\sqrt{1-x^2}}-\frac{3}{2}\right] ,\hspace{1.0cm}
\mbox{$\left(x < 1\right)$} \\ & \\
\frac{1}{(x^2-1)^2}\left[\frac{(2+x^2)\hspace{0.1cm}{\rm
arctan}\hspace{0.05cm}\sqrt{\frac{x-1}{1+x}}}
{\sqrt{x^2-1}}-\frac{3}{2}\right] ,\hspace{1.0cm}
\mbox{$\left(x > 1\right)$}
\end{array}
\right.
\end{equation}
The dimensionless surface mass density within $x$ is then,
\begin{equation}
\bar{\kappa}(x) = \frac{2}{x^2}\frac{M_{\infty}}{\pi r_s^2 \Sigma_{\rm{crit}_{\infty}}}
\int_0^x dx' \, x' \, f(x').
\end{equation}
The signal-to-noise ratio is then obtained by inserting the above relations
into equation (14).

As in the case of the NFW profile, given an overdensity of mass $M$,
there are three unknowns. We therefore apply the same energy
conservation argument as above, assuming the overdensity is initially
a uniform density sphere of radius $R_i$ and upon collapse relaxes to a
Hernquist profile. The potential energy upon collapse is
\begin{equation}
\Phi(y) = -\frac{G M_{\infty}^2}{6 r_s}
\left(1- \frac{6 y^2-8y+3}{y^4}\right),
\end{equation}
where $y=1+r/r_s$. Assuming energy conservation and a nearly
virialized overdensity yields $R=R_i=3.2 r_s$, allowing us to solve the
signal-to-noise relation.
\end{enumerate}
\label{lastpage}
\end{document}